\documentclass[11pt]{article}
\usepackage{jheppub}
\usepackage{amsmath,amssymb,graphicx}

\title{T-Duality Selection of a Disformal Effective Metric in String Gas Cosmology}

\author{Ali Nayeri}
\affiliation{Ordinal Research Institute, Wilmington, DE 19801, USA}
\emailAdd{ali.nayeri@ordinalresearch.org}


\abstract{We study a restricted disformal completion of string-frame string gas cosmology in which the light sector propagates on an effective metric with matter-cone opening $c(\phi)$. Requiring the gravitational coupling $A(\phi)=e^{-2\phi}c(\phi)^4$ to transform multiplicatively under T-duality reduces the problem to an exponential Cauchy equation, whose unique continuous solution is $c(\phi)=c_0 e^{-\alpha\phi}$. The coupling is therefore duality-selected, not assumed. Its magnitude $\alpha$ is UV-sensitive: a varying matter cone does not arise from the Weyl-invariant tree-level gauge action, but instead reflects higher-derivative and string-scale corrections. We then develop the structural consequences of this result. In the Einstein frame the dilaton remains healthy, with non-ghost kinetic coefficient $K(\alpha)=\frac{3}{2}(2+4\alpha)^2-4\ge 2$ for all $\alpha\ge0$. T-duality also defines a duality-invariant light-cone variable $\check c$, which provides a compact way to track the matter cone through duality-related branches and toward the self-dual point. As a controlled application on the analytic Hagedorn background, the wider early-time matter cone enlarges the comoving causal horizon by finite factors $1.15$ and $1.52$ for $\alpha=0.5$ and $1$, respectively, when the controlled phase is initialized at $t_i=0.02\,t_0$; for $\alpha\gtrsim1.15$, the cumulative horizon becomes dominated by the earliest times and correspondingly sensitive to the initial cutoff. The enhancement is therefore self-limiting through string-gas backreaction. Throughout, the construction is interpreted as an effective-metric statement in the controlled weak-coupling regime, not as a fundamental variation of the speed of light, with matching across the self-dual transition left as the main open problem.}

\begin{document}
\maketitle
\flushbottom

\section{Introduction}
%%%%%%%%%%%%%%%%%%%%%%%%%%%%%%%%%%%%%%%%%%%%%%%%%%%%%%%%%%%%

String gas cosmology (SGC) provides a noninflationary framework for the
early Universe in which the initial state is a quasi-static Hagedorn
phase~\cite{Hagedorn1965,AtickWitten88} populated by a thermal gas of
strings~\cite{Brandenberger1989,TseytlinVafa1992,Nayeri2005,Brandenberger2009,BW06};
for reviews see~\cite{Brandenberger2011,Brandenberger2023}.
Thermal string fluctuations generate an approximately scale-invariant
spectrum of scalar perturbations~\cite{Nayeri2005,Nayeri2006,BNPV2006more},
together with a characteristic blue-tilted spectrum of tensor modes from
closed-string thermodynamics~\cite{BNP2014}.
A natural question is whether the Hagedorn phase can also provide a
sufficiently large causal horizon without accelerated expansion.

In string frame, higher-derivative and string-scale corrections can
induce dilaton-dependent effective metrics for the light sector, so the
effective metric governing light-sector propagation is generically
\emph{disformally} related to the gravitational
metric~\cite{Bekenstein1993,BettoniLiberati2013,ZumalacarreguiGarciaBellido2014}.
Such field-dependent effective metrics allow the matter and
gravitational light cones to differ while each remains causal, with no
microscopic breaking of local Lorentz invariance. They also provide a
clean way to reformulate older varying-cone ideas in covariant effective
metric language~\cite{Moffat1993,Albrecht1999,Barrow1999,Magueijo2003}; for
more recent varying-light-speed cosmologies see~\cite{Moffat2016,Lee2023}.
The central result of this paper is structural. We do not derive the
disformal coupling from first principles; rather, we \emph{admit} a
restricted disformal completion in which the matter-cone opening is a
function $c(\phi)$ of the dilaton, and ask what T-duality requires of it.
Requiring the gravitational coupling $A(\phi)=e^{-2\phi}c(\phi)^4$ to
transform multiplicatively under T-duality reduces to an exponential
Cauchy equation whose unique continuous solution is
$c(\phi)=c_0\,e^{-\alpha\phi}$ (Sec.~\ref{sec:uniqueness}). Within this
class the exponential coupling is therefore \emph{duality-selected}, not
assumed; its magnitude $\alpha$ is UV-sensitive, since a varying matter
cone cannot arise from the Weyl-invariant tree-level gauge action and
instead reflects higher-derivative and string-scale
corrections~(Sec.~\ref{sec:origin}). We then develop the structural
consequences of this duality-selected metric: a healthy, non-ghost
Einstein frame, and a duality-invariant light-cone variable $\check c$
that tracks the matter cone across duality-related branches and toward
the self-dual point~(Sec.~\ref{sec:transition}). Only as a controlled
\emph{application}, on the analytic Hagedorn background, does the wider
early-time matter cone enlarge the comoving causal horizon by finite,
parameter-dependent factors during the weak-coupling Hagedorn phase.

The scope of the paper is deliberately limited. We work only within the
weak-coupling regime where the low-energy string-frame description is
under control, and we treat the approach to the self-dual point as the
boundary of validity of that description rather than as a controlled
transition. The matching across the self-dual regime is therefore left
open. T-duality nevertheless provides a nontrivial structural anchor: at
the self-dual radius $R=\ell_s$, the effective light cone is invariant,
making that point the natural place where any completion of the model
must restore the late-time relativistic branch. On the exponential
ansatz the effective cone has a non-monotonic profile with a universal
crossover time $t_{\rm cross}/t_0\approx 0.285$, a property of the
Hagedorn background itself rather than an independent model parameter.

%%%%%%%%%%%%%%%%%%%%%%%%%%%%%%%%%%%%%%%%%%%%%%%%%%%%%%%%%%%%
\section{Restricted Effective Theory}
%%%%%%%%%%%%%%%%%%%%%%%%%%%%%%%%%%%%%%%%%%%%%%%%%%%%%%%%%%%%

We begin from the string-frame action
\begin{equation}
\begin{aligned}
S ={}&
-\int d^4x\,\sqrt{-G}\,
\frac{1}{2\kappa(c)^2}e^{-2\phi}
\left(R+4\nabla_\mu\phi\nabla^\mu\phi\right) \\
&-\int d^4x\,\sqrt{-G}\,\rho
+\int d^4x\,\sqrt{-G}\,V(c,\phi),
\end{aligned}
\label{eq:action-original}
\end{equation}
with $\kappa(c)^2=8\pi G/c^4$ and $ds^2=-dt^2+e^{2\mu(t)}d\vec x^{\,2}$,
where $a(t)=e^{\mu(t)}$.
We restrict the theory by imposing $c=c(\phi)$ directly in the action.
Defining
\begin{equation}
A(\phi)\equiv e^{-2\phi}c(\phi)^4,
\qquad
V_{\rm eff}(\phi)\equiv V(c(\phi),\phi),
\label{eq:Aphi}
\end{equation}
The string-gas matter couples not to the gravitational metric
$G_{\mu\nu}$ but to a disformal matter metric~\cite{Bekenstein1993}
\begin{equation}
\hat G_{\mu\nu}=G_{\mu\nu}+\bigl(1-c(\phi)^2\bigr)\,u_\mu u_\nu,
\qquad
u_\mu\equiv-\frac{\partial_\mu\phi}
{\sqrt{-G^{\alpha\beta}\partial_\alpha\phi\,\partial_\beta\phi}},
\label{eq:disformal-metric}
\end{equation}
which on the homogeneous background ($\phi=\phi(t)$, $u_\mu=(1,0,0,0)$ in
the gauge $N=1$) reduces to
$d\hat s^2=-c(\phi)^2dt^2+a^2d\vec x^{\,2}$, so that light-sector signals
propagate with speed $c(\phi)/a$ while the gravitational background
remains canonical FRW. This is the minimal homogeneous disformal choice
realizing a $\phi$-dependent matter light cone; it assumes
$\partial_\mu\phi$ timelike, which holds on the background and for
modes below the gradient scale at which the timelike branch breaks down.
Throughout the controlled interval $c(\phi)^2=c_0^2e^{-2\alpha\phi}>0$ is
manifestly positive [Eq.~\eqref{eq:c-analytic}] and $\partial_\mu\phi$
stays timelike, so $\hat G_{\mu\nu}$ retains Lorentzian signature and the
matter equations of motion remain hyperbolic and well posed.
Because the matter sector is minimally coupled to $\hat G_{\mu\nu}$, it
obeys the exact matter-frame conservation law
$\hat\nabla_\mu\hat T^{\mu\nu}=0$, i.e.\
$\dot{\hat\rho}+3H(\hat\rho+\hat p)=0$ on the homogeneous background.
In Eqs.~\eqref{eq:minisuper}--\eqref{eq:phi-eqA}, $\rho$ denotes the
effective homogeneous density entering the gravity-frame minisuperspace
equations after projection from the matter-frame stress tensor.
With this understood, the minisuperspace Lagrangian for the FRW ansatz is
\begin{align}
L_{\rm mini}
&=
\frac{e^{3\mu}}{16\pi G}
\bigl[
6A\dot\mu^2
+6A_{,\phi}\dot\mu\dot\phi
+4A\dot\phi^2
-6kAe^{-2\mu}
\bigr] \nonumber\\
&\quad
-e^{3\mu}\rho
+e^{3\mu}V_{\rm eff}.
\label{eq:minisuper}
\end{align}

Varying with respect to the lapse, $\mu$, and $\phi$ gives,
respectively, the Hamiltonian constraint
\begin{equation}
3\dot\mu^2
+3\frac{A_{,\phi}}{A}\dot\mu\dot\phi
+2\dot\phi^2
+3ke^{-2\mu}
=
\frac{8\pi G}{A}\rho_{\rm tot},
\label{eq:constraintA}
\end{equation}
the $\mu$ equation
\begin{align}
2\ddot\mu+3\dot\mu^2
&+2\frac{A_{,\phi}}{A}\dot\mu\dot\phi
+\frac{A_{,\phi}}{A}\ddot\phi
+\frac{A_{,\phi\phi}}{A}\dot\phi^2 \nonumber\\
&-2\dot\phi^2
+ke^{-2\mu}
=
-\frac{8\pi G}{A}p_{\rm tot},
\label{eq:mu-eqA}
\end{align}
and the $\phi$ equation
\begin{align}
8A\,\ddot\phi
&+24A\,\dot\mu\dot\phi
+6A_{,\phi}\ddot\mu
+12A_{,\phi}\dot\mu^2 \nonumber\\
&+4A_{,\phi}\dot\phi^2
+6kA_{,\phi}e^{-2\mu}
+16\pi G\,\partial_\phi V_{\rm eff}
=0,
\label{eq:phi-eqA}
\end{align}
where $\rho_{\rm tot}=\rho+V_{\rm eff}$ and $p_{\rm tot}=p-V_{\rm eff}$.
Defining
\begin{equation}
C(\phi)\equiv \frac{d\ln c}{d\phi},
\qquad
\frac{A_{,\phi}}{A}=-2+4C(\phi),
\label{eq:Cphi}
\end{equation}
the Hamiltonian constraint becomes
\begin{equation}
3\dot\mu^2
-3\bigl(2-4C\bigr)\dot\mu\dot\phi
+2\dot\phi^2
+3ke^{-2\mu}
=
8\pi G\,e^{2\phi}c^{-4}\rho_{\rm tot}.
\label{eq:constraintC}
\end{equation}

In the limit $c(\phi)\to c_0$ (constant), $C\to 0$,
$A\to c_0^4 e^{-2\phi}$, and Eq.~\eqref{eq:constraintC} reduces to
\begin{equation}
3\dot\mu^2-6\dot\mu\dot\phi+2\dot\phi^2+3ke^{-2\mu}
=
8\pi G\,c_0^{-4}e^{2\phi}\rho_{\rm tot},
\label{eq:constraint-constant-c}
\end{equation}
which is the standard string-frame SGC background structure
in terms of the shifted dilaton $\varphi\equiv 2\phi-3\mu$~\cite{MeissnerVeneziano91}.
Thus the restricted theory smoothly recovers standard SGC once
the effective light cone freezes.

%%%%%%%%%%%%%%%%%%%%%%%%%%%%%%%%%%%%%%%%%%%%%%%%%%%%%%%%%%%%
\section{T-Duality Selection of the Disformal Coupling}
\label{sec:uniqueness}

In standard SGC the dilaton--gravity sector is invariant under T-duality
in the three large spatial dimensions, which acts on the minisuperspace
background as
\begin{equation}
\mu\to-\mu,\qquad \phi\to\phi-3\mu,
\label{eq:Tduality-fields}
\end{equation}
leaving the shifted dilaton $\varphi=2\phi-3\mu$
invariant~\cite{Veneziano1991,Giveon1994,Nayeri2006}. For $c=c_0$ the
gravitational coupling $A_0(\phi)=c_0^4\,e^{-2\phi}$ transforms as
$A_0(\phi-3\mu)=A_0(\phi)\,e^{6\mu}$, the factor $e^{6\mu}$ being
compensated by $\sqrt{-G}$ in the action so that the dynamics is duality
invariant. With a general $c(\phi)$ the coupling
$A(\phi)=e^{-2\phi}c(\phi)^4$ instead transforms as
\begin{equation}
A(\phi-3\mu)=e^{6\mu}\,e^{-2\phi}\,c(\phi-3\mu)^4.
\label{eq:A-Tduality}
\end{equation}

We require that $A$ transform \emph{multiplicatively},
$A(\phi-3\mu)=A(\phi)\,h(\mu)$. This is not a cosmetic assumption: under
$\phi\to\phi-3\mu$ the kinetic block of the minisuperspace
Lagrangian~\eqref{eq:minisuper} maps as
$L_{\rm kin}\to h(\mu)\,\tilde L_{\rm kin}$, with $\tilde L_{\rm kin}$ the
dual Lagrangian ($\mu\to-\mu$). If $A$ did not factor in this way, the
kinetic couplings $A_{,\phi}/A$ and $A_{,\phi\phi}/A$ in
Eqs.~\eqref{eq:constraintA}--\eqref{eq:phi-eqA} would acquire explicit
$\mu$-dependent corrections that no $\phi$-only field redefinition can
absorb, breaking the equivalence between the expanding and contracting
branches. Within the class of local redefinitions that preserve the
minisuperspace kinetic structure, multiplicativity is therefore the
natural requirement for the dual background to obey the same field
equations (up to the lapse rescaling); we do not exclude more exotic
completions with explicit $\mu$-dependence outside this class.

The requirement gives
$e^{6\mu}\,c(\phi-3\mu)^4=c(\phi)^4\,h(\mu)$. Setting $\phi=0$ fixes
$h(\mu)=e^{6\mu}\,c(-3\mu)^4/c_0^4$, and substituting back (taking the
positive fourth root) yields, for $f(x)\equiv c(x)/c_0$,
\begin{equation}
f(\phi-3\mu)=f(\phi)\,f(-3\mu),
\label{eq:cauchy}
\end{equation}
the exponential Cauchy equation. Its unique continuous solution is
$f(x)=e^{-\alpha x}$ for a constant $\alpha$, so that
\begin{equation}
\boxed{\;c(\phi)=c_0\,e^{-\alpha\phi},\qquad
h(\mu)=e^{(6+12\alpha)\mu}\;.}
\label{eq:unique-ansatz}
\end{equation}
The exponential form is thus \emph{not} an ad hoc ansatz but the unique
continuous disformal coupling for which the gravitational sector of the
string-frame action transforms multiplicatively under T-duality
(continuity excludes the pathological discontinuous solutions of the
Cauchy equation). The parameter $\alpha\geq 0$ measures the strength of
the coupling; $\alpha=0$ recovers standard SGC with $h=e^{6\mu}$.

\paragraph{Einstein-frame health.}
The same coupling defines a well-behaved Einstein frame. Under the
conformal rescaling $\tilde G_{\mu\nu}=A(\phi)\,G_{\mu\nu}$ with
$A=c_0^4\,e^{-(2+4\alpha)\phi}$ and $C\equiv A_{,\phi}/A=-(2+4\alpha)$,
the string-frame action reduces to
\begin{equation}
S_E=\frac{1}{16\pi G}\int d^4x\,\sqrt{-\tilde G}\,
\bigl[\tilde R-\mathcal{K}(\alpha)\,(\tilde\nabla\phi)^2\bigr],
\qquad
\mathcal{K}(\alpha)=\tfrac{3}{2}(2+4\alpha)^2-4.
\label{eq:einstein-frame}
\end{equation}
Since $\mathcal{K}(\alpha)\geq 2$ for all $\alpha\geq 0$, the dilaton
kinetic term retains the correct (non-ghost) sign: the duality-selected
coupling defines a healthy effective theory rather than a merely formal
ansatz, and at $\alpha=0$ it recovers the standard dilaton-gravity
normalization. The same conformal map clarifies the mechanism: the
Einstein-frame scale factor is $\tilde a=A^{1/2}a$, and since
$A\propto e^{-(2+4\alpha)\phi}$ the quasi-static string-frame Hagedorn
phase maps to an \emph{expanding}, non-accelerating Einstein frame whose
expansion rate increases with $\alpha$~\cite{BNPV2006more}. The enlarged
causal structure produced by the wide early-time matter cone in the
string frame is thus the same effect as the enhanced Einstein-frame
expansion.

\paragraph{Covariance of the potential.}
A dilaton potential $V_{\rm eff}(\phi)$ in the action must respect the
same symmetry. Under $\phi\to\phi-3\mu$ the potential term requires
$V_{\rm eff}(\phi-3\mu)=V_{\rm eff}(\phi)\,g(\mu)$, and the same Cauchy
argument fixes the unique continuous solution
$V_{\rm eff}(\phi)=V_0\,e^{-\sigma\phi}$. Here we set $V_{\rm eff}=0$,
which satisfies this trivially; a nonzero potential would add a
duality-constrained coupling $\sigma$ whose interplay with $\alpha$ is
left to future work.

%%%%%%%%%%%%%%%%%%%%%%%%%%%%%%%%%%%%%%%%%%%%%%%%%%%%%%%%%%%%
\section{Microscopic origin of the disformal coupling}
\label{sec:origin}

The duality argument of Sec.~\ref{sec:uniqueness} fixes the \emph{form} of
$c(\phi)$ but not the magnitude of $\alpha$. It is natural to ask whether
$\alpha$ is determined by how the light sector couples to the dilaton. A
tempting argument starts from the tree-level gauge kinetic term
$-\tfrac14 e^{-2\lambda\phi}F_{\mu\nu}F^{\mu\nu}$ and infers a
$\phi$-dependent photon speed $c\propto e^{-(\lambda-1)\phi}$, hence
$\alpha=\lambda-1$. This inference is incorrect: in four dimensions the
Maxwell action
$\sqrt{-G}\,G^{\mu\alpha}G^{\nu\beta}F_{\mu\nu}F_{\alpha\beta}$ is Weyl
invariant, so a scalar prefactor multiplies the gauge \emph{coupling} while
leaving the principal symbol---and hence the light cone---fixed by
$G_{\mu\nu}$. A dilaton-dependent coupling is not a dilaton-dependent light
cone.

A genuine matter-frame cone distinct from $G_{\mu\nu}$ requires a
\emph{derivative} (disformal) coupling of the light sector to
$\partial\phi$---the structure already encoded in
$\hat G_{\mu\nu}=G_{\mu\nu}+(1-c^2)u_\mu u_\nu$ with
$u_\mu\propto\partial_\mu\phi$ [Eq.~\eqref{eq:disformal-metric}],
equivalently an operator of the schematic form
$\partial_\mu\phi\,\partial_\nu\phi\,F^{\mu\alpha}F^{\nu}{}_{\alpha}$ that
breaks Weyl invariance along $\partial\phi$. Such terms are absent from the
two-derivative action but arise generically beyond it, with model-dependent
coefficients: (i)~from $\alpha'$ (higher-derivative) corrections of
Drummond--Hathrell type~\cite{DrummondHathrell1980}, in which vacuum
polarization in a nontrivial background shifts the photon cone; (ii)~from the
open-string (Seiberg--Witten) metric~\cite{SeibergWitten1999} seen by gauge
fields on a D-brane, which differs disformally from the closed-string metric
and becomes $\phi$-dependent in a varying-dilaton background; and (iii)~from
integrating out dilaton-dependent-mass Kaluza--Klein or winding modes, which
leaves $\phi$-dependent kinetic structures for the light fields. In each case
the coefficient---hence $\alpha$---is set by ultraviolet data (the
string/compactification scale, the brane configuration, the flux), not by the
low-energy symmetry.

The resulting picture is sharper than ``$\alpha$ is a free parameter.''
T-duality covariance fixes the functional \emph{form} of the disformal
coupling to be exponential (Sec.~\ref{sec:uniqueness}), while its
\emph{magnitude} $\alpha$ is the strength of a specific, duality-form-fixed
derivative operator whose coefficient is ultraviolet-sensitive and
model-dependent. A first-principles determination of $\alpha$ for a given
compactification---through the $\alpha'$-corrected or open-string effective
action---is a well-posed problem left to future work.

\section{Numerical Results}
%%%%%%%%%%%%%%%%%%%%%%%%%%%%%%%%%%%%%%%%%%%%%%%%%%%%%%%%%%%%

To make the disformal dynamics concrete, we adopt the duality-selected
coupling~\eqref{eq:unique-ansatz}
\begin{equation}
c(\phi) = c_0\,e^{-\alpha\phi},
\label{eq:exponential-ansatz}
\end{equation}
which gives $C(\phi) = -\alpha$ (constant) and $A_{,\phi}/A =
-(2+4\alpha)$. The parameter $\alpha \geq 0$ controls the disformal
coupling; $\alpha=0$ recovers standard SGC. Because this is the duality-selected coupling of
Sec.~\ref{sec:uniqueness}, it is not an ad hoc choice: it keeps
$c(\phi)$ positive, gives a constant $C(\phi)=-\alpha$, and is the
unique continuous form consistent with T-duality covariance of the
gravitational coupling. The present paper does not attempt to
derive $\alpha$ from a specific compactification or microscopic
completion; it is treated as an effective parameter of the disformal
sector. The benchmark values $\alpha\in\{0.5,1,2\}$ used below are
chosen only to illustrate the size of the controlled horizon
enhancement.

We initialize the background on the exact analytic Hagedorn solution
of Ref.~\cite{Nayeri2006}, Eqs.~(23)--(24):
\begin{equation}
\varphi(t) = \varphi_0 + \ln\!\left[
\frac{\mathcal{C}}{t\,(t_0-t)}\right],
\;\;
\mu(t) = \mu_0 + \mu_1\ln\!\left[\frac{t}{t_0-t}\right],
\label{eq:hagedorn-analytic}
\end{equation}
with $\mu_1 = 1/\sqrt{3}$ (expanding branch). The four-dimensional
dilaton $\phi$ is recovered from $\varphi = 2\phi - 3\mu$, and
initial conditions are read off at $t=t_i \ll t_0$. Explicitly, on the
expanding branch the four-dimensional dilaton is
\begin{equation}
\phi(t)=\phi_\ast+\tfrac{\sqrt3-1}{2}\ln t-\tfrac{\sqrt3+1}{2}\ln(t_0-t),
\label{eq:phi-analytic-expanding}
\end{equation}
and the duality-selected coupling takes the explicit power-law form
\begin{equation}
c(t)=c_\ast\,t^{-\alpha(\sqrt3-1)/2}\,(t_0-t)^{\alpha(\sqrt3+1)/2},
\qquad c_\ast\equiv c_0\,e^{-\alpha\phi_\ast},
\label{eq:c-analytic}
\end{equation}
confirming that the exponential coupling is natural for the logarithmic
dilaton evolution of the Hagedorn background. This structure in fact
generalizes to arbitrary $\alpha$: writing
\begin{equation}
\mu=\mu_0+m_e\ln t+m_\ell\ln(t_0-t),\quad
\phi=\phi_0+a_e\ln t+a_\ell\ln(t_0-t),
\label{eq:gen-log-ansatz}
\end{equation}
and substituting into the vacuum equations
\eqref{eq:mu-eqA}--\eqref{eq:phi-eqA} with $\rho=V_{\rm eff}=k=0$, the
early- and late-time coefficient pairs satisfy the \emph{same} algebraic
system (fixed by $B\equiv2+4\alpha$) while the cross terms cancel
identically, so~\eqref{eq:gen-log-ansatz} is an exact vacuum solution at
every $\alpha$. The asymptotic ratios $r\equiv\dot\phi/\dot\mu$ are the
roots of
\begin{equation}
2r^2-3(2+4\alpha)\,r+3=0,
\label{eq:r-constraint}
\end{equation}
recovering $r_0=(3+\sqrt3)/2$ at $\alpha=0$. Because the Hagedorn phase
keeps its quasi-static logarithmic structure at all $\alpha$, the
standard SGC thermal-fluctuation formalism~\cite{BNPV2006more} applies on
the generalized background without modification.

\begin{figure}[t]
\centering
\includegraphics[width=0.98\columnwidth]{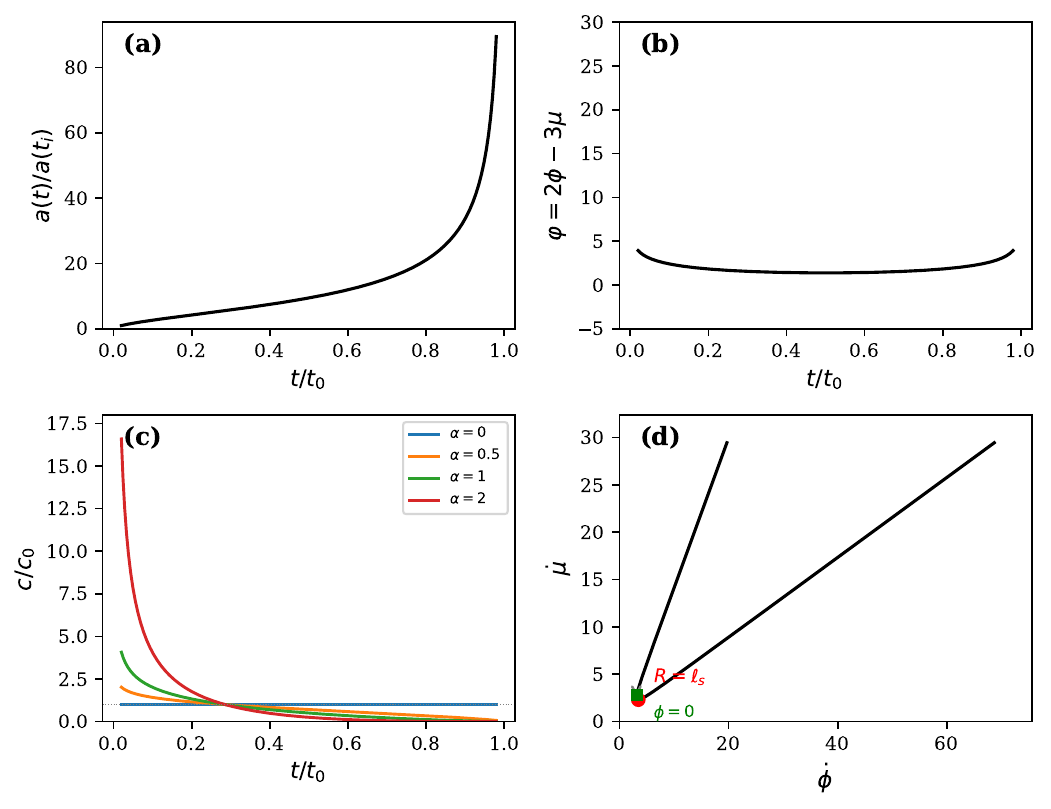}
\caption{(a) Scale factor $a(t)/a(t_i)$. (b) Shifted dilaton
$\varphi=2\phi-3\mu$. (c) Propagation speed $c/c_0$.
(d) Phase portrait $(\dot\phi,\dot\mu)$. Ansatz
$c(\phi)=c_0 e^{-\alpha\phi}$, $k=0$, $V_{\rm eff}=0$.
Solid: analytic~\eqref{eq:hagedorn-analytic}; dashed: numeric.}
\label{fig:background}
\end{figure}

We integrate Eqs.~\eqref{eq:mu-eqA}--\eqref{eq:phi-eqA} with $k=0$
and $V_{\rm eff}=0$ using an eighth-order Dormand--Prince scheme with
relative tolerance $10^{-11}$. Figure~\ref{fig:background} shows the
scale factor, shifted dilaton, propagation speed, and phase portrait
for $\alpha \in \{0, 0.5, 1, 2\}$. Solid curves are the analytic
background~\eqref{eq:hagedorn-analytic}; dashed curves are the
numerical solutions initialized thereon. The close overlap across all
panels confirms internal consistency of the equations of motion. The
shifted dilaton diverges as $t \to t_0$ for all $\alpha$, marking the
breakdown of the weak-coupling description at the self-dual point.

%%%%%%%%%%%%%%%%%%%%%%%%%%%%%%%%%%%%%%%%%%%%%%%%%%%%%%%%%%%%
\section{Dilaton-Induced Disformal Light-Cone Profile and T-Duality}
%%%%%%%%%%%%%%%%%%%%%%%%%%%%%%%%%%%%%%%%%%%%%%%%%%%%%%%%%%%%

\begin{figure}[t]
\centering
\includegraphics[width=0.98\textwidth]{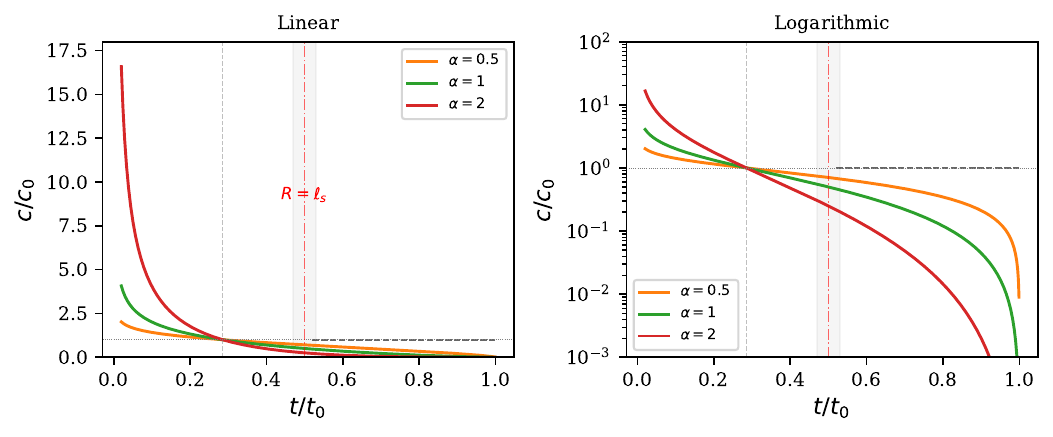}
\caption{The disformal light-cone profile on linear (left) and logarithmic
(right) scales. Solid: weak-coupling phase; dashed: illustrative
late-time relativistic branch $c=c_0$; grey band: strong-coupling transition; dash-dotted red:
self-dual point $R=\ell_s$ ($\mu=0$, $t=t_0/2$). Crossover at the
universal time $t_{\rm cross}/t_0\approx 0.285$, independent of $\alpha$.}
\label{fig:bounce}
\end{figure}

The numerical solutions exhibit a non-monotonic disformal light-cone
profile. For $\alpha>0$ the dilaton starts negative and grows
monotonically, so the matter light cone is initially wider than the
gravitational one ($c>c_0$), crosses $c=c_0$ when $\phi=0$, and then
narrows as the self-dual regime is approached. The crossover occurs
where $\phi=0$, i.e.\ $\varphi=-3\mu$, giving the universal time
\begin{equation}
\frac{t_{\rm cross}}{t_0} \approx 0.285,
\label{eq:tcross}
\end{equation}
independently of $\alpha$, so it is a property of the Hagedorn
background rather than a tunable parameter of the disformal ansatz.
This behavior is shown in Fig.~\ref{fig:bounce} on both linear and
logarithmic scales. We emphasize that the early wide-cone regime
($c>c_0$) is a statement about the matter light cone relative to the
gravitational one in the disformal frame, not superluminal signaling or
a microscopic breaking of local Lorentz invariance; each sector
propagates causally on its own cone.

The formal late-time behavior $c\to 0$ as $t\to t_0$ should not be
over-interpreted. It occurs precisely where the shifted dilaton signals
the breakdown of the weak-coupling description, so at most it suggests
that the light sector becomes strongly coupled near the self-dual
transition. We therefore use this behavior only as an indicator of
where the effective theory ceases to be reliable, not as a controlled
prediction of the transition itself.

The self-dual point $R=\ell_s$ ($\mu=0$, $t=t_0/2$) is anchored by
T-duality. Under T-duality in the three large dimensions~\cite{TseytlinVafa1992,Giveon1994,Nayeri2006},
\begin{equation}
\mu \to -\mu, \qquad \phi \to \phi - 3\mu,
\label{eq:Tduality}
\end{equation}
the shifted dilaton $\varphi = 2\phi - 3\mu$ is invariant~\cite{Nayeri2006}.
For $c(\phi) = c_0\,e^{-\alpha\phi}$, this gives
\begin{equation}
c \;\to\; c\,e^{3\alpha\mu}.
\label{eq:c-Tduality}
\end{equation}
Three consequences follow. First, at $\mu=0$, $c$ is T-duality
invariant, so the self-dual point provides a natural anchor for the
normalization of the effective light cone. Second, on the expanding
branch ($\mu > 0$), T-duality maps to the contracting branch
($\mu < 0$) with a wider matter cone, consistent with the expected
symmetry. Third, although the present weak-coupling analysis does not
derive the post-transition matching, it identifies the self-dual point
as the place where any completion must restore the late-time
relativistic branch.

%%%%%%%%%%%%%%%%%%%%%%%%%%%%%%%%%%%%%%%%%%%%%%%%%%%%%%%%%%%%
\section{Flatness and Horizon Growth}
%%%%%%%%%%%%%%%%%%%%%%%%%%%%%%%%%%%%%%%%%%%%%%%%%%%%%%%%%%%%

In the quasi-static Hagedorn phase, the comoving particle horizon is
\begin{equation}
d_H(t)=\int_{t_i}^t \frac{c(t')}{a(t')}dt'.
\label{eq:horizon}
\end{equation}
Since $a(t)$ is approximately constant, a wider early-time matter cone
substantially enlarges the causal horizon, providing contact over
scales that would otherwise remain disconnected without accelerated
expansion.

\begin{figure}[t]
\centering
\includegraphics[width=0.98\textwidth]{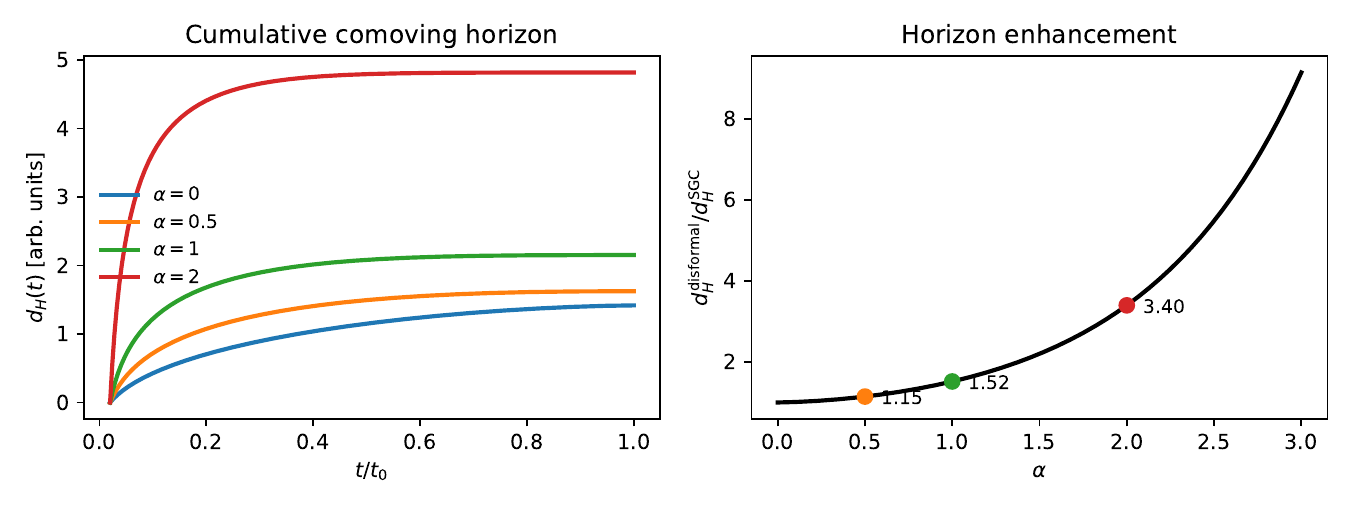}
\caption{Left: cumulative comoving horizon $d_H(t)$ on the analytic
Hagedorn background for $\alpha\in\{0,0.5,1,2\}$. Right: horizon
enhancement ratio $d_H^{\rm disformal}/d_H^{\rm SGC}$ as a function of
$\alpha$, evaluated with cutoff $t_i=0.02\,t_0$. For
$\alpha\gtrsim\alpha_c\approx1.15$ the ratio is dominated by the lower
cutoff (steep rise); the convergent regime ($\alpha<\alpha_c$, with a finite $t_i\to0$
limit) is $\alpha\lesssim1$, of which only $\alpha=0.5$ is weakly
cutoff-sensitive.}
\label{fig:horizon}
\end{figure}

We evaluate~\eqref{eq:horizon} on the analytic
background~\eqref{eq:hagedorn-analytic}, integrating from the onset of
the controlled phase at $t_i=0.02\,t_0$, using the same cutoff for all
$\alpha$ so that the comparison is at fixed initial data. One caveat must
be stated up front. Near the initial time the integrand behaves as
$c/a\sim t^{-[\alpha(\sqrt3-1)/2+1/\sqrt3]}$, which is integrable as
$t_i\to0$ only for
$\alpha<\alpha_c\equiv(2-2/\sqrt3)/(\sqrt3-1)\approx1.15$. For
$\alpha<\alpha_c$ the enhancement therefore has a well-defined
$t_i\to0$ limit, though the approach is slow near the threshold so
finite-$t_i$ values retain some residual sensitivity; for
$\alpha\gtrsim\alpha_c$ the cumulative horizon is dominated by the
earliest times and the ratio depends entirely on the choice of $t_i$. Figure~\ref{fig:horizon}
shows the cumulative horizon $d_H(t)$ and the enhancement ratio
$d_H^{\rm disformal}/d_H^{\rm SGC}$, which grows monotonically with
$\alpha$:
\begin{equation}
\frac{d_H^{\rm disformal}}{d_H^{\rm SGC}}\bigg|_{\alpha=0.5} \approx 1.15,
\;\;
\bigg|_{\alpha=1} \approx 1.52,
\;\;
\bigg|_{\alpha=2} \approx 3.40\ \ (\alpha>\alpha_c).
\label{eq:horizon-ratios}
\end{equation}
These span three regimes. At $\alpha=0.5$ the ratio is convergent and
only mildly cutoff-sensitive (it shifts by $\sim\!30\%$ as $t_i$ is
lowered from $0.02\,t_0$ to $10^{-4}t_0$). At $\alpha=1$, which sits just
below $\alpha_c$, the integral still converges as $t_i\to0$ but slowly,
so the finite-$t_i$ value retains moderate cutoff sensitivity (a factor
$\sim\!2$ over the same range). At $\alpha=2>\alpha_c$ the integral
\emph{diverges} as $t_i\to0$: the value is set entirely by the lower
cutoff and is quoted only to illustrate the trend, not as a
cutoff-independent prediction. The controlled statement is thus the
$\alpha\lesssim1$ enhancement at the stated $t_i=0.02\,t_0$. The dominant contribution comes from the early wide-cone
regime.
For a Hagedorn duration $T_H$, $d_H\sim c(t_i,\alpha)\,T_H$, so
the disformal coupling \emph{parametrically relaxes the required Hagedorn
duration} by $e^{|\alpha\phi(t_i)|}$. The robust factors ($1.15\times$, $1.52\times$ for $\alpha=0.5,1$) are
not competing with inflation-scale expansion, but rather reduce the
fine-tuning on how long the Hagedorn phase must persist. More explicitly, with
$c(t_i)=c_0\,e^{-\alpha\phi(t_i)}$ and $\phi(t_i)<0$ in the early
weak-coupling phase, the enhancement is \emph{exponential} in the
initial weak-coupling depth,
\begin{equation}
\frac{d_H^{\rm disformal}}{d_H^{\rm SGC}}\;\sim\;e^{\alpha|\phi(t_i)|},
\label{eq:horizon-parametric}
\end{equation}
so the tabulated factors~\eqref{eq:horizon-ratios} correspond to a
\emph{modest} initial displacement $|\phi(t_i)|=\mathcal{O}(1)$, while a
cosmologically significant enhancement is available at larger
$|\phi(t_i)|$, i.e.\ deeper initial weak coupling. As discussed below,
this enhancement is not arbitrarily tunable: the same trajectory must
remain within the controlled weak-coupling regime, which bounds $\alpha$
from above. This is a concrete, parameter-dependent
effect of the restricted disformal SGC framework. Since $c(\phi)$ is a
dilaton-dependent scalar in string frame, the varying light cone
reflects the field-dependent effective metric of the string sector
rather than a spontaneous breaking of Lorentz invariance at the
microscopic level.

\paragraph{Analytic flatness exponent.}
The flatness parameter $\epsilon\equiv\Omega-1\propto c^2/(a^2H^2)$ can be
evaluated in closed form on the analytic background. Using
\eqref{eq:hagedorn-analytic} and the explicit coupling~\eqref{eq:c-analytic},
with $H=\dot\mu=(t_0/\sqrt3)/[t(t_0-t)]$ and $a=(t/(t_0-t))^{1/\sqrt3}$,
\begin{equation}
\epsilon(t)\propto t^{\beta(\alpha)}(t_0-t)^{\gamma(\alpha)},\quad
\beta=2-\tfrac{2}{\sqrt3}-\alpha(\sqrt3-1),\;\;
\gamma=2+\tfrac{2}{\sqrt3}+\alpha(1+\sqrt3).
\label{eq:gamma}
\end{equation}
The curvature dilution on approach to the self-dual point is governed by
$\gamma(\alpha)$, which grows linearly with the disformal coupling,
$d\gamma/d\alpha=1+\sqrt3\approx2.73$ (with $\gamma_0=2+2/\sqrt3\approx3.15$
at $\alpha=0$). The coupling thus makes the suppression rate
\emph{parametrically controllable}; consistent with the caveat below, the
exponent~\eqref{eq:gamma} measures what the controlled weak-coupling phase
\emph{can} achieve and thereby bounds the work left to the transition.

Following the effective-metric flatness logic often associated with
varying-cone models~\cite{Albrecht1999,Barrow1999,BarrowMagueijo1999},
one defines $\epsilon\equiv\Omega-1 =
-kc^2/(a^2H^2)$: a rapid decrease in $c$ suppresses the curvature
term relative to the critical density. In the present model this
mechanism operates \emph{dynamically} from the ansatz. As the self-dual
regime is approached within the controlled interval, $c$ decreases while
$(aH)^2$ grows; numerically, $c^2/(aH)^2$ decreases with a log-derivative
that is negative and accelerating toward $t_0$.
Over $t/t_0\in[0.85,0.95]$, the suppression factor decreases by
$3.6\times10^{-4}$ ($\alpha=1$), $1.9\times10^{-5}$ ($\alpha=2$),
and $9.9\times10^{-7}$ ($\alpha=3$). Thus $\Omega-1$ is driven toward
zero within the controlled regime, at a rate growing with $\alpha$. We stress, however, that the strongest suppression accrues as
$t\to t_0$---i.e.\ as the weak-coupling description itself degrades---so
the firm, controlled statement is the milder suppression over earlier,
fully controlled times; the $t/t_0\in[0.85,0.95]$ values are indicative
of the trend rather than an established resolution of the flatness
problem.
Whether it fully resolves flatness here depends on physics through the
self-dual transition~\cite{Magueijo2009}.

\paragraph{Backreaction and the usable range of $\alpha$.}
The background used above is the dilaton-vacuum solution
($\rho=0$, $V_{\rm eff}=0$). This is self-consistent precisely in the
early weak-coupling phase that generates the horizon enhancement: the
string-gas source enters the Hamiltonian
constraint~\eqref{eq:constraintC}. Treating $\rho_{\rm tot}$ as
approximately constant over the quasi-static interval for this scaling
estimate, one has
$8\pi G\,e^{2\phi}c^{-4}\rho_{\rm tot}
\sim (8\pi G/c_0^4)\,e^{(2+4\alpha)\phi}\rho_{\rm tot}$,
which is exponentially \emph{suppressed} as $\phi\to-\infty$
($t\to t_i$) and grows monotonically toward the self-dual point
($\phi\to+\infty$). Backreaction of the string gas is therefore a
second, independent reason---alongside the divergence of $\varphi$---that
the late-time regime lies outside the controlled description. Since the
source steepens with the coupling (rate $2+4\alpha$), larger $\alpha$
shrinks the interval over which the vacuum background is reliable, so the
usable enhancement $e^{\alpha|\phi(t_i)|}$ is bounded above rather than
arbitrarily tunable. A rough estimate of the closing scale follows from
comparing the gas source to the dilaton-kinetic term that dominates the
vacuum constraint. Writing the fractional backreaction as
$B(\phi)\equiv 8\pi G\,e^{2\phi}c^{-4}\rho_{\rm tot}/\dot\phi^{2}
\propto e^{(2+4\alpha)\phi}$---treating $\rho_{\rm tot}$ as approximately constant in
the quasi-static Hagedorn phase and $\dot\phi^{2}$ varying only as a power
of $t$---backreaction reaches $\mathcal{O}(1)$ near
\begin{equation}
\phi_\ast \simeq \frac{1}{2+4\alpha}\,
\ln\!\Big(\frac{c_0^{4}\,\dot\phi^{2}}{8\pi G\,\rho_{\rm tot}}\Big).
\label{eq:phistar}
\end{equation}
The controlled window thus closes at $\phi_\ast$, which scales as
$1/(2+4\alpha)$, whereas the gain per unit dilaton interval grows only as
$\alpha$; the net usable enhancement $e^{\alpha(\phi_\ast-\phi_i)}$
therefore saturates rather than increasing without bound in $\alpha$.
A sharp bound requires integrating the coupled
background-plus-gas system and is left to future work; the point here is
that $\phi_\ast$ acts as a safety boundary delimiting the interval over
which the vacuum trajectories used above remain trustworthy, so the
mechanism is self-limiting.

%%%%%%%%%%%%%%%%%%%%%%%%%%%%%%%%%%%%%%%%%%%%%%%%%%%%%%%%%%%%
\section{Thermal Fluctuations and Controlled Regime}
%%%%%%%%%%%%%%%%%%%%%%%%%%%%%%%%%%%%%%%%%%%%%%%%%%%%%%%%%%%%

The nearly scale-invariant scalar spectrum of SGC is generated
thermally during the quasi-static Hagedorn
phase~\cite{Nayeri2005,Nayeri2006,BNPV2006more,BW06}. The disformal framework modifies the
causal structure of that phase without replacing its thermodynamic
origin: the wider early-time matter cone kinematically enlarges the horizon
and broadens the range of scales in thermal contact
(Eq.~\eqref{eq:horizon-ratios}). The controlled regime ends as
$\varphi\to\infty$ at $t\to t_0$, so the transition to the late-time
relativistic branch lies outside the present weak-coupling treatment.
How the disformal structure modifies the perturbation spectrum through
that transition is left for future work. We also delimit the scope of the horizon result: what is enlarged
here is the comoving horizon of the \emph{early} Hagedorn phase. Relating
this to the observed horizon problem---causal contact across the
last-scattering surface---requires propagating the enhancement through
the self-dual transition into the radiation era, which lies beyond the
present treatment. The controlled claim is thus the early-phase
enhancement itself, not a complete resolution of the horizon problem; a
complementary, fully duality-covariant approach to the size and horizon
problems in string cosmology has been developed in~\cite{BBF2020}.

At leading order the thermal string gas should still be formulated in the
matter frame defined by the disformal metric~\eqref{eq:disformal-metric},
with local stress tensor
\begin{equation}
\hat T_{\mu\nu}
=
(\hat\rho+\hat p)\hat u_\mu \hat u_\nu
+ \hat p\,\hat G_{\mu\nu}
+ \hat\Pi_{\mu\nu}.
\end{equation}
The crossover at $c(\varphi)=c_0$ is then not itself a thermodynamic
transition but a kinematic coincidence of the matter and gravitational
cones. Accordingly, the local Hagedorn equation of state and thermal
correlators are expected to remain smooth through $t_{\rm cross}$ in the
matter frame. What changes is the projection of $\hat T_{\mu\nu}$ into the
gravity-frame equations, where the disformal relation induces
dilaton-dependent exchange terms in the effective scalar source seen by
metric perturbations. The crossover is thus smooth for the local thermal
fluid; its cosmological import lies in the changing map between
matter-frame thermodynamics and gravity-frame causal structure.

%%%%%%%%%%%%%%%%%%%%%%%%%%%%%%%%%%%%%%%%%%%%%%%%%%%%%%%%%%%%
\section{Duality constraints on the self-dual transition}
\label{sec:transition}

The matching across the self-dual point lies outside the weak-coupling
description used here. T-duality nevertheless constrains any completion,
independently of the non-perturbative dynamics.

Under $\mu\to-\mu$, $\phi\to\phi-3\mu$ [Eq.~\eqref{eq:Tduality}] the cone
transforms as $c\to c\,e^{3\alpha\mu}$ [Eq.~\eqref{eq:c-Tduality}] while the
shifted dilaton $\varphi=2\phi-3\mu$ is invariant. These combine into a
duality-\emph{invariant} light-cone variable
\begin{equation}
\check c\equiv c\,e^{3\alpha\mu/2}=c_0\,e^{-\alpha\varphi/2},
\label{eq:invariant-cone}
\end{equation}
which depends only on the invariant clock $\varphi$ and satisfies
$\check c=c$ at the self-dual point $\mu=0$.

Equation~\eqref{eq:invariant-cone} turns the matching into a constraint: a
duality-covariant completion must evolve the single invariant
$\check c(\varphi)$ and cannot let the matching depend on the duality-odd
variable $\mu$. Since the pre-transition (weak-coupling, $\mu>0$) and
post-transition (large-$R$, radiation) branches are exchanged by
$\mu\to-\mu$, they must join through the $\mu=0$ fixed point at a common
value of $\check c$. In particular, any non-perturbative completion---worldsheet
or matrix-string---that preserves T-duality must, on exiting the
high-density phase, track this single invariant $\check c(\varphi)$ rather
than $c$ and $\mu$ separately, as in the duality-covariant cosmologies of
double field theory~\cite{DFTcosmo1,DFTcosmo2}.

\paragraph{Conjectural extrapolation.}
The next step lies outside the controlled regime and is offered as a
duality-motivated expectation, not a result. If the non-perturbative
completion preserves T-duality and keeps $\check c(\varphi)$ continuous
through the self-dual region, then as $\varphi$ relaxes to its
post-Hagedorn value the invariant cone $\check c=c_0\,e^{-\alpha\varphi/2}$
approaches $c_0$; since $c=\check c$ on the late-time relativistic branch,
the observable light cone is driven to $c\to c_0$. Restoration of the
standard relativistic cone after the transition would then \emph{follow}
from duality covariance and continuity rather than being assumed.
Establishing these conditions requires the non-perturbative physics at
$R=\ell_s$ (e.g.\ a worldsheet or matrix-string treatment) and remains
open; what duality supplies unconditionally is the invariant
variable~\eqref{eq:invariant-cone} and the prohibition of $\mu$-dependent
matching.

\section{Discussion}
%%%%%%%%%%%%%%%%%%%%%%%%%%%%%%%%%%%%%%%%%%%%%%%%%%%%%%%%%%%%

The controlled result of the present analysis is modest and kinematical:
in string-frame SGC, a dilaton-dependent disformal matter metric widens
the matter light cone relative to the gravitational one during the early
weak-coupling Hagedorn phase and thereby enlarges the cumulative comoving
horizon by finite, parameter-dependent factors. For $\alpha=0.5$ and
$\alpha=1$ the convergent enhancement factors are $1.15$ and $1.52$ (at
$t_i=0.02\,t_0$, with mild and moderate residual cutoff sensitivity
respectively); larger $\alpha$ enters a cutoff-dominated regime and is
not quoted as a controlled number. The curvature contribution $c^2/(a^2H^2)$ is
correspondingly reduced over the late controlled interval.

T-duality gives this construction a nontrivial structural anchor. At the
self-dual radius the effective light cone is invariant under the duality
map, so the self-dual point is the natural place at which any completion
of the model must address the transition back to a standard relativistic
branch. What the present calculation does \emph{not} provide is that
transition itself. The weak-coupling description breaks down as the
self-dual regime is approached, and the formal late-time narrowing of the
matter cone should therefore be interpreted only as a marker of the
boundary of validity.

The claim of the paper is accordingly limited. We do not claim a
fundamental variation of the speed of light, and we do not claim a
controlled bounce solution through the self-dual transition. The result is
instead that string-frame SGC naturally contains a disformal
effective-metric structure whose controlled early-time effect is to relax
the causal-horizon requirements of the Hagedorn phase. Extending the
analysis through the self-dual region will require a non-perturbative
treatment or an independent dynamical matching mechanism. A complementary,
fully local T-duality-covariant formulation of the Hagedorn phase remains
an interesting direction for future work.

%%%%%%%%%%%%%%%%%%%%%%%%%%%%%%%%%%%%%%%%%%%%%%%%%%%%%%%%%%%%
\acknowledgments
I thank Robert Brandenberger for valuable discussions and for
comments on the manuscript.

%%%%%%%%%%%%%%%%%%%%%%%%%%%%%%%%%%%%%%%%%%%%%%%%%%%%%%%%%%%%

\section*{Declarations}
\textbf{Funding.} No funding was received for this work.
\textbf{Conflict of interest.} The author declares that he has no
competing interests.
\textbf{Data availability.} No data were generated or analysed; this is a
theoretical study.


\begin{thebibliography}{99}

\bibitem{Bekenstein1993}
J.~D.~Bekenstein,
Phys.\ Rev.\ D \textbf{48}, 3641 (1993).

\bibitem{BettoniLiberati2013}
D.~Bettoni and S.~Liberati,
Phys.\ Rev.\ D \textbf{88}, 084020 (2013).

\bibitem{ZumalacarreguiGarciaBellido2014}
M.~Zumalac\'arregui and J.~Garc\'ia-Bellido,
Phys.\ Rev.\ D \textbf{89}, 064046 (2014).

\bibitem{Hagedorn1965}
R.~Hagedorn,
Nuovo Cim.\ Suppl.\ \textbf{3}, 147 (1965).

\bibitem{AtickWitten88}
J.~J.~Atick and E.~Witten,
Nucl.\ Phys.\ B \textbf{310}, 291 (1988).

\bibitem{Brandenberger1989}
R.~H.~Brandenberger and C.~Vafa,
Nucl.\ Phys.\ B \textbf{316}, 391 (1989).

\bibitem{TseytlinVafa1992}
A.~A.~Tseytlin and C.~Vafa,
Nucl.\ Phys.\ B \textbf{372}, 443 (1992).

\bibitem{Brandenberger2009}
R.~H.~Brandenberger,
in \emph{String Cosmology}, ed.\ J.~Erdmenger (Wiley, 2009);
arXiv:0808.0746.

\bibitem{BW06}
T.~Battefeld and S.~Watson,
Rev.\ Mod.\ Phys.\ \textbf{78}, 435 (2006).

\bibitem{Nayeri2005}
A.~Nayeri, R.~Brandenberger and C.~Vafa,
Phys.\ Rev.\ Lett.\ \textbf{97}, 021302 (2006).

\bibitem{Nayeri2006}
A.~Nayeri, arXiv:hep-th/0607073.

\bibitem{BNPV2006more}
R.~H.~Brandenberger, A.~Nayeri, S.~P.~Patil and C.~Vafa,
Int.\ J.\ Mod.\ Phys.\ A \textbf{22}, 3621 (2007).

\bibitem{Moffat1993}
J.~W.~Moffat,
Int.\ J.\ Mod.\ Phys.\ D \textbf{2}, 351 (1993).

\bibitem{Albrecht1999}
A.~Albrecht and J.~Magueijo,
Phys.\ Rev.\ D \textbf{59}, 043516 (1999).

\bibitem{Magueijo2003}
J.~Magueijo,
Rep.\ Prog.\ Phys.\ \textbf{66}, 2025 (2003).

\bibitem{Barrow1999}
J.~D.~Barrow,
Phys.\ Rev.\ D \textbf{59}, 043515 (1999).

\bibitem{BarrowMagueijo1999}
J.~D.~Barrow and J.~Magueijo,
Phys.\ Lett.\ B \textbf{447}, 246 (1999).

\bibitem{Magueijo2009}
J.~Magueijo,
Phys.\ Rev.\ D \textbf{79}, 043525 (2009).

\bibitem{MeissnerVeneziano91}
K.~A.~Meissner and G.~Veneziano,
Phys.\ Lett.\ B \textbf{267}, 33 (1991).

\bibitem{Veneziano1991}
G.~Veneziano,
Phys.\ Lett.\ B \textbf{265}, 287 (1991).

\bibitem{Giveon1994}
A.~Giveon, M.~Porrati and E.~Rabinovici,
Phys.\ Rep.\ \textbf{244}, 77 (1994).

\bibitem{DrummondHathrell1980}
I.~T.~Drummond and S.~J.~Hathrell,
Phys.\ Rev.\ D \textbf{22}, 343 (1980).

\bibitem{SeibergWitten1999}
N.~Seiberg and E.~Witten,
JHEP \textbf{09}, 032 (1999) [hep-th/9908142].

\bibitem{Brandenberger2011}
R.~H.~Brandenberger,
Class.\ Quantum Grav.\ \textbf{28}, 204005 (2011) [arXiv:1105.3247].

\bibitem{DFTcosmo1}
R.~Brandenberger, R.~Costa, G.~Franzmann and A.~Weltman,
Phys.\ Rev.\ D \textbf{99}, 023531 (2019) [arXiv:1809.03482].

\bibitem{DFTcosmo2}
H.~Bernardo, R.~Brandenberger and G.~Franzmann,
Phys.\ Rev.\ D \textbf{99}, 063521 (2019) [arXiv:1901.01209].

\bibitem{BBF2020}
H.~Bernardo, R.~Brandenberger and G.~Franzmann,
JHEP \textbf{10}, 155 (2020) [arXiv:2007.14096].

\bibitem{Brandenberger2023}
R.~Brandenberger,
JCAP \textbf{11}, 019 (2023) [arXiv:2306.12458].

\bibitem{BNP2014}
R.~H.~Brandenberger, A.~Nayeri and S.~P.~Patil,
Phys.\ Rev.\ D \textbf{90}, 067301 (2014) [arXiv:1403.4927].

\bibitem{Moffat2016}
J.~W.~Moffat,
Eur.\ Phys.\ J.\ C \textbf{76}, 130 (2016) [arXiv:1404.5567].

\bibitem{Lee2023}
S.~Lee,
Found.\ Phys.\ \textbf{53}, 40 (2023) [arXiv:2303.13772].

\end{thebibliography}
\end{document}